# The asymmetrical Acquisition of information about the range of asset value in market

Jianhao Su
sdwhfxsx@163.com
Shandong Normal university

Yanliang Zhang
zhyanliang@sina.com
Shandong University of Finance and Economics

## Abstract

The information investors acquire in asset markets has various forms. We refer to range information as information about the upper and lower bound which the payoff of an asset may reach in the future. This paper explores the market impacts of investors' asymmetrical acquisition of range information. Uninformed traders are inherently unable to directly obtain the private signal held by informed traders. This study shows that when range information is released to investors asymmetrically, uninformed traders who can only obtain rougher range information will not trade assets under the max-min ambiguity aversion criterion. Investors' asymmetrical acquisition of range information can cause that market liquidity and the sensitivity of market price to private signal vary continuously with the signal and noise trading volume. We also reveal that investors' asymmetrical acquisition of range information can increase market liquidity and the sensitivity of price under some conditions and decrease them under some other conditions.

## 1. Introduction

In financial market, the completeness of information held by different traders varies, which is known as information asymmetry among traders. The information distribution among market investors has a great impact on the operation of the market. A large number of literature specifically focuses on the impact of information distribution on the formation of market prices, the reaction of market prices to released signals and market quality. Market quality mainly includes indicators such as market liquidity, price volatility and market efficiency.

A large part of the theoretical research on this topic is developed under the framework of rational expectation. This series of studies assumes that traders with less information can use market price to infer the private information held by traders who have more information, and apply their inference to make investment decisions.

Grossman and Stigliz(1980) is an early study based on the rational expectations framework. In its setup, there are informed traders and uninformed traders in the market. Informed traders can obtain a private signal about asset value, while uninformed traders can only observe asset price and then infer the private signal of informed traders through the price. Under this setup, Grossman and Stigliz(1980) analyzed the trading behavior of the two types of traders and market equilibrium.

In recent years, research literature on this topic has focused more on investors' acquisition and learning of information, as well as the impact of their information acquisition behavior on market prices and market quality. Also based on the framework of rational expectation model, this series of literature argues that the process of investors learning information can be interfered by some factors and become more complex, which lowers the effectiveness of investors information learning. These factors mainly include:

(1) the ambiguity in investors' cognition of some parameters about the assets;

(2) the situation where asset value is composed of multiple fundamentals;

(3) deviation in investors' interpretation of the information in market;

Since these factors can damage the effectiveness of investors' information learning, thereby affecting their trading behavior, market price formation, and market quality.

The first type of the studies is developed under the assumption that traders' belief about some key parameters is ambiguous, which makes it more complicated for uninformed traders to infer the private information of informed traders from market price, thereby making their trading behavior more conservative. For example, Epstein and Schneider(2008) discussed the situation where investors have ambiguity in their perception of the accuracy of signal; Mele and Sangiorgi(2015), Hahn and Kwon(2015), and Condie and Ganguli(2017) focused on the situation where there is ambiguity in investors' belief about the expected value of an asset; Easley et al.(2014), Huang et al.(2017),

Illeditsch et al.(2021) considered the situation where investors are ambiguous about the correlation coefficient between different asset values. These studies show that the ambiguity of investors can lead to significant changes in their trading volume, market prices and market quality, such as the sensitivity of market price, asset premium, market liquidity.

The second type of the studies analyzes the market where asset value is determined by multiple fundamentals, and different types of informed traders can obtain information about different fundamentals, such as Kondor(2012), Goldtein and Yang(2015), Liu et al.(2019), Yang and Zhu(2020). When various traders participate in transactions, the information about multiple fundamentals will be injected into market price simultaneously and these fundamentals will be reflected in the price in a complicated form, which can make it more difficult for investors to infer the information from the price and reducing the accuracy of their inference results. The above literature shows that investors' trading behavior, the market price, and the market quality will be distinct given that the value of asset is determined by multiple fundamentals.

The third type of the studies is developed under the assumption that there is a deviation in investors' interpretation of the information in market, which can make their trading decision deviate from the optimal. They analyzes how market equilibrium price and market quality would be affected by the deviation, such as Dudams et al.(2017), Banerjee et al.(2018), Atmaz and Basak(2018), and Hu and Wang(2024).

Related literature usually depicts the information investors acquire in two ways. The asset has an uncertain future value $v$. Based on the assumption that $v$ consists of several parts, such as $v = v_1 + v_2$, the first way is to take the realized value of $v_1$ or $v_2$ as the information acquired by investors.

Use $\eta$ to represent a stochastic error item. The second way is to take the realized value of $v + \eta$ as the information acquired by investors.

The two ways of depicting the information investors receive are essentially the same and we refer to them as common-form signals. For reflecting the information asymmetry in market, related literature usually assumes that informed traders can receive this form of signal while uninformed traders cannot, or informed traders can receive more signals than uninformed traders.

However, there are more forms of information that investors can obtain in the financial market. Some information can reflect the upper or lower bound of the future value of assets. We refer to this type of information as the range information about asset value.

For example, many countries have a price limit regime in their securities markets, under which investors believe that the future value of securities must have an upper and lower bound.

Additionally, different security analysts often make different predictions about the future value of a company's stocks. Optimistic analysts tend to give a higher predicted value, while pessimistic analysts tend to give a lower one. Some investors will regard the lowest predicted value given by analysts as the lower bound of the stock's future value, and the highest as the upper bound of its future value. In other words, the stock's future value will be between the two in these investors' belief.

In some cases, a company or some financial institutions will take some powerful measures to stop the downward trend of its price under the condition that the stock price of the company falls to a certain level. In similar cases, some blockholders may sell their shares in large quantities to realize profits when the stock price rises to a certain level, which can prevent the price from rising further. Once the insider information above are acquired by some investors, they will also form the belief about the upper and lower bound of the asset value in the future.

To sum up, range information is likely to exist in investors' information set and affects their belief. The main issue we want to explore is the impact of investors' acquisition of range information beyond common-form signals on their trading behavior, the formation of market price, and market quality. This study helps to improve our understanding of the operation of financial markets and enhance the ability of existing literature to explain phenomena in financial markets.

For introducing information asymmetry between investors, we make reference to related literature, assuming that informed traders can receive a private signal in common-form while uninformed traders cannot obtain it. Given that the information acquired by informed traders inclines to be more accurate than that acquired by uninformed traders, we also assume that the range information acquired by uninformed traders is rougher. In other words, we further depict the information asymmetry between investors through the heterogeneity of their range information, which has not appeared in existing literature.

In our main model, the acquisition of range information makes the composition of informed traders' information set more complicated, and the asymmetry of the range information acquired by traders also makes it more difficult for uninformed traders to infer the private signal of informed traders through the price. Therefore, our work adds to the related literature based on the rational expectations framework which focuses more on investors' complex acquisition of information.

The remainder of this paper is organized as follows. Section 2 presents our main model by which we analyze the trading behavior of various traders and the market equilibrium in the situation where traders acquire range information with asymmetry. In section 3, we use a benchmark model to present the trading behavior of various traders and the market equilibrium in the situation where traders do not receive any range information. Section 4 compares the equilibrium price, the sensitivity of the price to private signal, market liquidity, and market efficiency under the two models to reveal the impact of traders' asymmetric acquisition of range information. Finally, section 5 concludes.

## 2. The Main Model

### 2.1 Setup

The asset market lasts for two periods: t=0 and 1. Investors trade assets at t=0 and get the payoffs of the assets they hold at t=1.

There are two assets traded in the market. The first asset is a risk-free bond which is unlimited supply. Referring to the setup of Easley et al.(2014) and Mondria et al.(2022), we also suppose that its payoff at t=1 and its price at t=0 are both 1 for simplicity. The second asset is a risky asset (such as stock of a company) which has a total supply of $Z$ units.

Use $v$ to denote the payoff of the risky asset at t=1. At t=0, $v$ is uncertain for all investors and can be regarded as the value of the risky asset in the future.

According to Grossman and Stiglitz(1980), Huang et al.(2020), we assume that

$$v = u + \varepsilon,$$

where $u \sim N(\mu_0, \sigma_u^2)$ and $\varepsilon \sim N(0, \sigma_\varepsilon^2)$. $u$ and $\varepsilon$ are mutually independent. $u$ can be regarded as the fundamental of the company and $\varepsilon$ can be referred to as the disturbance of some unobservable random factors on its future value.

At t=0, investors can finance their purchase of risky assets by selling risk-free bonds short, and there are no restrictions on the amount of short selling.

Following the setup of Goldstein and Yang(2017,2022), there are three types of investors in the market: informed traders, uninformed traders and noise traders. Suppose the total mass of the first two types of investors is 1. The fraction of informed traders and uniformed traders are respectively $x_I$ and $x_U$ with $x_I + x_U = 1$.

Noise traders totally demand $y$ units of the risky asset at t=0, that is, $y$ is noise trading volume. $y \sim N(0, \sigma_y^2)$ and $y$ is independent of $u$ and $\varepsilon$. A larger $\sigma_y^2$ implies a greater size of noise trading in the market. In the following, we use $\tilde{y}$ to denote the realized value of $y$.

At t=0, informed traders can observe the realized value of $u$ while uninformed traders cannot and $\varepsilon$ is unobservable to all investors. So $u$ can be regarded as informed traders' private signal about $v$. A larger value of its variance $\sigma_u^2$ implies a greater informativeness of the private signal. $\varepsilon$ should be treated as the error of informed traders' private signal $u$. Given that $Var[v|u=\tilde{u}]=\sigma_\varepsilon^2$, a smaller value of its variance $\sigma_\varepsilon^2$ implies a smaller deviation of $u$ from the asset future value $v$ and then a more precise signal received by informed traders.

Additionally, we suppose that investors can acquire range information about $v$ at t=0. The range information acquired by informed traders implies the upper and lower bound of the asset future value are $\overline{v}$ and $\underline{v}$ respectively, that is, they form the belief $v \in [\underline{v}, \overline{v}]$. To further reflect the information asymmetry between investors, we assume that uninformed traders can only acquire rougher range information and form the belief $v \in [\underline{v}_0, \overline{v}_0]$, where $\underline{v}_0 < \underline{v}$ and $\overline{v} < \overline{v}_0$, that is, $[\underline{v}, \overline{v}]$is a proper subset of $[\underline{v}_0, \overline{v}_0]$. This means the value range perceived by informed traders is more precise than that perceived by uninformed traders. We can take $\overline{v}_0=+\infty$ or $\underline{v}_0=-\infty$ to represent the case where uninformed traders don't acquire any information about the upper or lower bound of $v$ respectively.

Related literature typically assumes that uninformed traders are aware that informed traders' information are more sufficient than theirs, such as Mele and Sangiorgi(2015), Easley and O'Hara(2009), Grossman and Stigliz(1980). We adhere to this assumption by supposing that uninformed traders perceive that their range information is rougher than informed traders'.

Before trading, every investor is endowed with $D_0$ units of the risk-free bond.

At t=0, the price of the risky asset is denoted by $p$. Use $\theta_I$ and $\theta_U$ to denote informed traders' and uninformed traders' demand for the risky asset respectively, so their wealth at t=1 are respectively

$$W_I = D_0 + \theta_I(v-p)$$

and

$$W_U = D_0 + \theta_U(v-p).$$

Suppose that all the investors have a CARA utility function with a common absolute risk aversion coefficient $\gamma$, that is,

$$U(W) = -e^{-\gamma \cdot W},$$

where $W$ is the wealth held by the investor at t=1. At t=0, every investor trade in order to maximize the expectation of her utility conditional on her information set.

Under our assumptions, the market clearing condition should be

$$x_I\theta_I + x_U\theta_U + \tilde{y} = Z. \qquad (2-1)$$

2.2 The decision of informed traders

If the price of the risky asset $p \geq \overline{v}$, selling the asset short will make a profit with probability 1 and then informed traders are bound to do that infinitely. If $p \leq \underline{v}$, informed traders are bound to buy the asset infinitely. So only when $\underline{v} < p < \overline{v}$ can the market reach equilibrium. The premise of our following discussion is $p \in (\underline{v}, \overline{v})$.

Now consider a representative informed trader. The realized value of $u$ is denoted by $\tilde{u}$, and his information set is $\{u = \tilde{u},\ v \in [\underline{v},\ \overline{v}]\}$. The conditional expectation of his utility is

$$U_I(\theta_I) \triangleq E\left[-e^{-\gamma(D_0 + \theta_I(v-p))} \middle| u = \tilde{u}, v \in [\underline{v}, \overline{v}]\right].$$

His decision problem is

$$\max_{\theta_I}\ U_I(\theta_I).$$

Use $\Psi(\cdot)$ and $\psi(\cdot)$ to denote the distribution and density function of the standard normal distribution respectively. We can derive

$$U_I(\theta_I) = \frac{E\left[-e^{-\gamma(D_0+\theta_I(v-p))}\mathbf{1}_{\{v\in[\underline{v},\overline{v}]\}} | u = \tilde{u}\right]}{E\left[\mathbf{1}_{\{v\in[\underline{v},\overline{v}]\}} | u = \tilde{u}\right]}$$

$$= -e^{-\gamma D_0 + \gamma(p-\tilde{u})\theta_I + \frac{\gamma^2\sigma_\varepsilon^2\theta_I^2}{2}} \cdot \frac{\Psi\left(\frac{\overline{v} + \gamma\sigma_\varepsilon^2\theta_I - \tilde{u}}{\sigma_\varepsilon}\right) - \Psi\left(\frac{\underline{v} + \gamma\sigma_\varepsilon^2\theta_I - \tilde{u}}{\sigma_\varepsilon}\right)}{\Psi\left(\frac{\overline{v} - \tilde{u}}{\sigma_\varepsilon}\right) - \Psi\left(\frac{\underline{v} - \tilde{u}}{\sigma_\varepsilon}\right)}. \qquad (2-2)$$

We give the details about the calculation of the equation above in appendix A.1. The derivative of the utility with respect to $\theta_I$ is

$$U_I'(\theta_I) = -\gamma \cdot U_I(\theta_I) \cdot \left[\boldsymbol{J}_{[\underline{v},\ \overline{v}]}(\tilde{u} - \gamma\sigma_\varepsilon^2\theta_I) - p\right],$$

where $\boldsymbol{J}_{[\underline{v},\ \overline{v}]}(t) \triangleq \sigma_\varepsilon \left[\frac{\psi\left(\frac{\underline{v}-t}{\sigma_\varepsilon}\right) - \psi\left(\frac{\overline{v}-t}{\sigma_\varepsilon}\right)}{\Psi\left(\frac{\overline{v}-t}{\sigma_\varepsilon}\right) - \Psi\left(\frac{\underline{v}-t}{\sigma_\varepsilon}\right)}\right] + t.$

It can be proved that there exists a unique $\widetilde{\theta_I}$ such that $U_I'(\widetilde{\theta_I}) =$

0 and $U_I(\theta_I)$ reaches the maximum at $\theta_I = \widetilde{\theta_I}$. To define $L(\theta_I) \triangleq \boldsymbol{J}_{[\underline{v},\ \overline{v}]}(\tilde{u} - \gamma\sigma_\varepsilon^2\theta_I) - p$, it follows that $U_I{}'(\theta_I) = -\gamma \cdot U_I(\theta_I) \cdot L(\theta_I)$. According to the item (3) in appendix A.2, we have

$$L'(\theta_I) = -\gamma\sigma_\varepsilon^2 \cdot \boldsymbol{J}_{[\underline{v},\ \overline{v}]}{}'(\tilde{u} - \gamma\sigma_\varepsilon^2\theta_I) < 0,$$

that is, $L(\theta_I)$ is a strictly monotone decreasing function for $\theta_I \in R$. Given $\underline{v} < p < \overline{v}$ and the item (4) in appendix A.2, we have

$$\lim_{\theta_I \to -\infty} L(\theta_I) = \overline{v} - p > 0 \text{ and } \lim_{\theta_I \to +\infty} L(\theta_I) = \underline{v} - p < 0.$$

In addition, considering the continuity of $L(\theta_I)$, $L(\theta_I) = 0$ definitely has a unique solution $\widetilde{\theta_I}$.

For $\theta_I < \widetilde{\theta_I}$, we have $L(\theta_I) > 0$ and then $U_I{}'(\theta_I) = -\gamma \cdot U_I(\theta_I) \cdot L(\theta_I) > 0$. For $\theta_I > \widetilde{\theta_I}$, we have $L(\theta_I) < 0$ and then $U_I{}'(\theta_I) < 0$. So $\widetilde{\theta_I}$ can maximize the utility $U_I(\theta_I)$ and satisfies

$$\boldsymbol{J}_{[\underline{v},\ \overline{v}]}\left(\tilde{u} - \gamma\sigma_\varepsilon^2\widetilde{\theta_I}\right) - p = 0. \tag{2-3}$$

We give a closed form solution of an informed trader's optimal decision by the following proposition.

**Proposition 1.** For a given price $p$, if $p \in (\underline{v},\ \overline{v})$, an informed trader's optimal demand for the risky asset is $\widetilde{\theta_I} = \frac{1}{\gamma\sigma_\varepsilon^2} \cdot \tilde{u} + k$, where the parameter $k$ is the solution of

$$p = \boldsymbol{J}_{[\underline{v},\ \overline{v}]}(-\gamma\sigma_\varepsilon^2 k). \tag{2-4}$$

The equation above has a unique solution. $\widetilde{\theta_I}$ decreases monotonically with the price $p$.

According the item (3) of appendix A.2, we have $\boldsymbol{J}_{[\underline{v},\ \overline{v}]}{}'(t) > 0$, which implies $\boldsymbol{J}_{[\underline{v},\ \overline{v}]}(t)$ is continuous and increases monotonically with $t$. So it must have an inverse function $\boldsymbol{J}^{-1}_{[\underline{v},\ \overline{v}]}(\cdot)$. Proposition 1 actually shows that the optimal demand is

$$\widetilde{\theta_I} = \frac{1}{\gamma\sigma_\varepsilon^2} \cdot \tilde{u} - \frac{\boldsymbol{J}^{-1}_{[\underline{v},\ \overline{v}]}(p)}{\gamma\sigma_\varepsilon^2},$$

where $\widetilde{\theta_I}$ includes two parts. The first part $\frac{1}{\gamma\sigma_\varepsilon^2}\cdot\tilde{u}$ is determined by the private signal $\tilde{u}$ and cannot be affected by the range information(i.e., the lower bound $\underline{v}$ and the upper bound $\overline{v}$) and the price $p$. On the contrary, the second part $-\frac{J^{-1}_{[\underline{v},\ \overline{v}]}(p)}{\gamma\sigma_\varepsilon^2}$ is determined by the range information and the price $p$, but cannot be affected by the private signal $\tilde{u}$.

## 2.3 The decision of uninformed traders

Uniformed traders cannot receive the private signal $\tilde{u}$, but can observe the price $p$ and a rougher information range $[\underline{v}_0, \overline{v}_0]$.

Informed traders inject their private signal into the price when they trade in the market. Thus, related literature under the framework of rational expectation usually assumes that uninformed traders try to infer the private signal from the price. In our model, the price is also impacted by the range information $[\underline{v},\overline{v}]$ because informed traders' trading volume is also decided by $\underline{v}$ and $\overline{v}$. As a result, uninformed traders' unacquaintance about $[\underline{v},\overline{v}]$ can make their inference about the private signal more difficult.

Before analyzing uninformed traders' inference, related literature always conjectures the equilibrium price $p$, setting $p$ in a linear form, such as Easley et al.(2014), Goldstein and Yang(2017). However, under our setup, $p$ must be between $\underline{v}$ and $\overline{v}$, which implies that it cannot be set in a linear form. According to equation (2-4), we conjecture the equilibrium price as a quasi-linear form

$$p = \boldsymbol{J}_{[\underline{v},\ \overline{v}]}(\tau\tilde{u} + \alpha\tilde{y} + \beta), \tag{2-5}$$

where the coefficients $\tau, \alpha, \beta$ will be determined in equilibrium. According to the item (1) of appendix A.2, if $p$ is given by equation(2-5), it follows that $p \in (\underline{v}, \overline{v})$.

According to Proposition 1, under the price given by equation(2-5), we have $k = -\frac{\tau}{\gamma\sigma_\varepsilon^2}\tilde{u} - \frac{\alpha}{\gamma\sigma_\varepsilon^2}\tilde{y} - \frac{\beta}{\gamma\sigma_\varepsilon^2}$ and an informed trader' demand is

$$\widetilde{\theta_I} = \frac{1-\tau}{\gamma\sigma_\varepsilon^2}\tilde{u} - \frac{\alpha}{\gamma\sigma_\varepsilon^2}\tilde{y} - \frac{\beta}{\gamma\sigma_\varepsilon^2}. \tag{2-6}$$

Remind that uninformed traders perceive the range information

$[\underline{v}_0, \overline{v}_0]$ they acquire is rougher than informed traders' in our setup. Use $[\underline{v}', \overline{v}']$ to denote the range information acquired by informed traders in the belief of uninformed traders. $\overline{v}'$ and $\underline{v}'$ definitely satisfy $\underline{v}_0 < \underline{v}' < p < \overline{v}' < \overline{v}_0$. As for uninformed traders, the reason why $\underline{v}' < p < \overline{v}'$ holds is only under this condition can the market realize equilibrium. It can be seen that they face ambiguity about $\overline{v}'$ and $\underline{v}'$ actually.

Define $\Omega_p \triangleq \{[\underline{v}', \overline{v}'] \mid \underline{v}' \in (\underline{v}_0, p), \overline{v}' \in (p, \overline{v}_0)\}$. In the belief of uninformed traders, the more precise range information acquired by informed traders must belong to $\Omega_p$. Now consider a representative uninformed trader. For an arbitrary member $[\underline{v}', \overline{v}'] \in \Omega_p$, if he considers it as informed traders' range information, according to our conjecture about the price (i.e., equation (2-5)), he will take the price as

$$p = \boldsymbol{J}_{[\underline{v}', \overline{v}']}(\tau\tilde{u} + \alpha\tilde{y} + \beta).$$

Based on this, he can infer that

$$\tilde{u} + \frac{\alpha}{\tau}\tilde{y} = \frac{\boldsymbol{J}^{-1}_{[\underline{v}', \overline{v}']}(p) - \beta}{\tau}.$$

It can be seen that his inference about the private signal $\tilde{u}$ is polluted by the noise trading $\tilde{y}$ and varies with $\underline{v}'$ and $\overline{v}'$.

In this case, the conditional expectation of his utility at t=1 should be

$$U_U(\theta_U; \underline{v}', \overline{v}') \triangleq E\left[-e^{-\gamma(D_0 + \theta_U(v-p))} \middle| \tilde{u} + \frac{\alpha}{\tau}\tilde{y} = \frac{\boldsymbol{J}^{-1}_{[\underline{v}', \overline{v}']}(p) - \beta}{\tau}, v \in [\underline{v}', \overline{v}']\right].$$

By some calculations, we derive

$$U_U(\theta_U; \underline{v}', \overline{v}') = e^{\gamma(p-\mu_\eta)\theta_U + \frac{\gamma^2\sigma_\eta^2\theta_U^2}{2}} \cdot \left[\frac{\Psi\left(\frac{\overline{v}' + \gamma\sigma_\eta^2\theta_U - \mu_\eta}{\sigma_\eta}\right) - \Psi\left(\frac{\underline{v}' + \gamma\sigma_\eta^2\theta_U - \mu_\eta}{\sigma_\eta}\right)}{\Psi\left(\frac{\overline{v}' - \mu_\eta}{\sigma_\eta}\right) - \Psi\left(\frac{\underline{v}' - \mu_\eta}{\sigma_\eta}\right)}\right], \qquad (2-7)$$

where $\mu_\eta = \omega_1\mu_0 + \omega_2\left(\frac{J^{-1}_{[\underline{v}',\ \overline{v}']}(p)-\beta}{\tau}\right), \sigma_\eta^2 = \sigma_\varepsilon^2 + \omega_1\sigma_u^2$ with $\omega_1 = \frac{\alpha^2\sigma_y^2}{\tau^2\sigma_u^2+\alpha^2\sigma_y^2}, \omega_2 = 1-\omega_1 = \frac{\tau^2\sigma_u^2}{\tau^2\sigma_u^2+\alpha^2\sigma_y^2}$. The details about the calculation of the equation above can be seen in appendix A.4.

As for the optimal decision of the uninformed trader, we have the following lemma. The proof is given in appendix A.5.

**Lemma 1.** For any $\underline{v}' \in (\underline{v}_0, p)$ and $\overline{v}' \in (p, \overline{v}_0)$, $U_U(\theta_U; \underline{v}', \overline{v}')$ reaches its maximum at

$$\bar{\theta}_U = \frac{1}{\gamma\sigma_\eta^2}\left[\omega_1\mu_0 - \frac{\omega_2\beta}{\tau} + \left(\frac{\omega_2}{\tau} - 1\right)\cdot J^{-1}_{[\underline{v}',\overline{v}']}(p)\right].$$

Additionally, $U_U(\theta_U; \underline{v}', \overline{v}')$ increases monotonically in $(-\infty, \bar{\theta}_U)$ and decreases monotonically in $(\bar{\theta}_U, +\infty)$.

It can be proved that the coefficient of $J^{-1}_{[\underline{v}',\overline{v}']}(p)$ is not zero, that is, $\frac{\omega_2}{\tau} - 1 \neq 0$.

Suppose $\tau = \omega_2$. Then the uninformed trader's optimal demand for the risky asset is $\bar{\theta}_U = \frac{1}{\gamma\sigma_\eta^2}\left[\omega_1\mu_0 - \frac{\omega_2\beta}{\tau}\right]$. Substituting this expression and the informed trader's optimal demand given by equation (2-6) into the market clearing condition(equation (2-1)), we derive

$$\frac{x_I(1-\tau)}{\gamma\sigma_\varepsilon^2}\tilde{u} + \left(1 - \frac{\alpha x_I}{\gamma\sigma_\varepsilon^2}\right)\tilde{y} - \frac{\beta x_I}{\gamma\sigma_\varepsilon^2} + \frac{x_U}{\gamma\sigma_\eta^2}\left[\omega_1\mu_0 - \frac{\omega_2\beta}{\tau}\right] = Z.$$

Comparing the coefficient of $\tilde{u}$ on the left-hand side with that on the right-hand side, we have $\frac{x_I(1-\tau)}{\gamma\sigma_\varepsilon^2} = 0$ and $1 - \frac{\alpha x_I}{\gamma\sigma_\varepsilon^2} = 0$, that is, $\tau = 1$ and $\alpha = \frac{\gamma\sigma_\varepsilon^2}{x_I} > 0$. It follows that $\omega_2 = 1$. However, we have $\omega_2 = \frac{\tau^2\sigma_u^2}{\tau^2\sigma_u^2+\alpha^2\sigma_y^2} < 1$, which is inconsistent with the case $\omega_2 = 1$. Thus, $\tau \neq \omega_2$ and then $\frac{\omega_2}{\tau} - 1 \neq 0$.

Since $\frac{\omega_2}{\tau} - 1 \neq 0$, Lemma 1 suggests that the uninformed trader' optimal decision $\bar{\theta}_U$ varies with $\underline{v}'$ and $\overline{v}'$, whereas he is ambiguous about the two parameters. In order to determine his trading volume, we comply with the max-min ambiguity aversion criterion which is adopted by Gilboa and Schmeidler(1989), Gajdos et al.(2008), Easley et al.(2014), etc.

Under this criterion, he aims to maximize

$$U_U(\theta_U) \triangleq \min_{\substack{\underline{v}' \in (\underline{v}_0, p) \\ \overline{v}' \in (p, \overline{v}_0)}} U_U(\theta_U; \underline{v}', \overline{v}').$$

His decision problem is

$$\max_{\theta_U} U_U(\theta_U) = \max_{\theta_U} \min_{\substack{\underline{v}' \in (\underline{v}_0, p) \\ \overline{v}' \in (p, \overline{v}_0)}} U_U(\theta_U; \underline{v}', \overline{v}').$$

For solving this problem, we first give the following lemma and its proof can be seen in appendix A.6.

**Lemma 2.** For any fixed $\underline{v}' \in (\underline{v}_0, p)$, $\lim_{\delta \to 0+} \boldsymbol{J}^{-1}_{[\underline{v}',\ p+\delta]}(p) = +\infty$. For any fixed $\overline{v}' \in (p, \overline{v}_0)$, $\lim_{\delta \to 0+} \boldsymbol{J}^{-1}_{[p-\delta,\ \overline{v}']}(p) = -\infty$.

According to equation (2-7), for any given $\underline{v}' \in (\underline{v}_0, p)$ and $\overline{v}' \in (p, \overline{v}_0)$, we have $U_U(0; \underline{v}', \overline{v}') = -1$. It follows that $U_U(0) = -1$.

Since $\frac{\omega_2}{\tau} - 1 \neq 0$, we will discuss under the case of $\frac{\omega_2}{\tau} - 1 > 0$ and $\frac{\omega_2}{\tau} - 1 < 0$ respectively.

Suppose $\frac{\omega_2}{\tau} - 1 > 0$. For any fixed $\overline{v}'' \in (p, \overline{v}_0)$, Lemma 2 implies that there exists a small positive $\delta$ such that

$$\boldsymbol{J}^{-1}_{[p-\delta,\ \overline{v}'']}(p) < -\left(\omega_1 \mu_0 - \frac{\omega_2 \beta}{\tau}\right) \Big/ \left(\frac{\omega_2}{\tau} - 1\right).$$

Take $\bar{\theta}_U{}' = \frac{1}{\gamma \sigma_\eta^2}\left[\omega_1 \mu_0 - \frac{\omega_2 \beta}{\tau} + \left(\frac{\omega_2}{\tau} - 1\right) \cdot \boldsymbol{J}^{-1}_{[p-\delta,\ \overline{v}'']}(p)\right]$ and we can derive $\bar{\theta}_U{}' < 0$ by the inequality above. According to Lemma 1, $U_U(\cdot\,; p-\delta, \overline{v}'')$ decreases monotonically in $(\bar{\theta}_U{}', +\infty)$. Thus, for any $\theta_U > 0$, we have

$$U_U(\theta_U; p-\delta, \overline{v}'') < U_U(0; p-\delta, \overline{v}'') = -1,$$

which implies that for the uninformed trader, not participating in the trading of the risky asset is better than buying assets if he thinks the lower bound of the asset value is very closed to the price $p$. For any $\theta_U > 0$, by the definition of $U_U(\cdot)$, it follows that

$$U_U(\theta_U) \leq U_U(\theta_U; p-\delta, \overline{v}'') < -1.$$

On the other hand, for any fixed $\underline{v}'' \in (\underline{v}_0, p)$, Lemma 2 implies that there exists a small positive $\delta$ such that

$$\boldsymbol{J}^{-1}_{[\underline{v}'',\ p+\delta]}(p) > -\left(\omega_1\mu_0 - \frac{\omega_2\beta}{\tau}\right)\Big/\left(\frac{\omega_2}{\tau} - 1\right).$$

Take $\bar{\theta}_U'' = \frac{1}{\gamma\sigma_\eta^2}\left[\omega_1\mu_0 - \frac{\omega_2\beta}{\tau} + \left(\frac{\omega_2}{\tau} - 1\right)\cdot\boldsymbol{J}^{-1}_{[\underline{v}'',\ p+\delta]}(p)\right]$ and we can derive $\bar{\theta}_U'' > 0$. According to lemma 1, $U_U(\cdot\,; \underline{v}'', p+\delta)$ increases monotonically in $(-\infty, \bar{\theta}_U'')$. Thus, for any $\theta_U < 0$, we have

$$U_U(\theta_U; \underline{v}'', p+\delta) < U_U(0; \underline{v}'', p+\delta) = -1,$$

which implies that for the uninformed trader, not participating in the trading of the risky asset is better than short selling assets if he considers the upper bound of the asset value is very closed to the price $p$. For any $\theta_U < 0$, by the definition of $U_U(\cdot)$, it follows that

$$U_U(\theta_U) \le U_U(\theta_U; \underline{v}'', p+\delta) < -1.$$

In summary, under the case of $\frac{\omega_2}{\tau} - 1 > 0$, we manifest that for any $\theta_U \neq 0$,

$$U_U(\theta_U) < -1 = U_U(0),$$

which implies an uninformed trader' s optimal trading volume is 0, that is, giving up trading is his best decision.

Under the case of $\frac{\omega_2}{\tau} - 1 < 0$, for any $\theta_U \neq 0$, we have

$$U_U(\theta_U) < -1 = U_U(0)$$

holds for any $\theta_U \neq 0$. This can be demonstrated in a similar way as above. Then, we derive the following proposition.

**Proposition 2.** If the range information $[\underline{v}_0, \bar{v}_0]$ acquired by uninformed traders is rougher than that acquired by informed traders, the optimal trading volume of an uninformed trader is $\widetilde{\theta_U} = 0$ under the max-min ambiguity aversion criterion.

2.4 The market equilibrium

In the market equilibrium, the demand of an uninformed trader is $\widetilde{\theta_U} = 0$, and the demand of an informed trader is specified as equation (2-6).

Insert their trading volume into the market clearing condition (i.e., equation (2-1)), we derive

$$\frac{x_I(1-\tau)}{\gamma\sigma_\varepsilon^2}\tilde{u}+\left(1-\frac{\alpha x_I}{\gamma\sigma_\varepsilon^2}\right)\tilde{y}-\frac{\beta x_I}{\gamma\sigma_\varepsilon^2}=Z.$$

Comparing the coefficient of $\tilde{u}$, $\tilde{y}$ and the intercept item on the left-hand side with that on the right-hand side, we obtain

$$\tau=1,\qquad \alpha=\frac{\gamma\sigma_\varepsilon^2}{x_I},\qquad \beta=-\frac{Z\gamma\sigma_\varepsilon^2}{x_I}.$$

Therefore, the equilibrium price is

$$p_1=\boldsymbol{J}_{[\underline{v},\ \overline{v}]}\left(\tilde{u}+\frac{\gamma\sigma_\varepsilon^2}{x_I}\tilde{y}-\frac{Z\gamma\sigma_\varepsilon^2}{x_I}\right). \tag{2-8}$$

More concretely,

$$p_1=\sigma_\varepsilon\cdot\left[\frac{\psi\left(\frac{\underline{v}-\left(\tilde{u}+\frac{\gamma\sigma_\varepsilon^2}{x_I}\tilde{y}-\frac{Z\gamma\sigma_\varepsilon^2}{x_I}\right)}{\sigma_\varepsilon}\right)-\psi\left(\frac{\overline{v}-\left(\tilde{u}+\frac{\gamma\sigma_\varepsilon^2}{x_I}\tilde{y}-\frac{Z\gamma\sigma_\varepsilon^2}{x_I}\right)}{\sigma_\varepsilon}\right)}{\Psi\left(\frac{\overline{v}-\left(\tilde{u}+\frac{\gamma\sigma_\varepsilon^2}{x_I}\tilde{y}-\frac{Z\gamma\sigma_\varepsilon^2}{x_I}\right)}{\sigma_\varepsilon}\right)-\Psi\left(\frac{\underline{v}-\left(\tilde{u}+\frac{\gamma\sigma_\varepsilon^2}{x_I}\tilde{y}-\frac{Z\gamma\sigma_\varepsilon^2}{x_I}\right)}{\sigma_\varepsilon}\right)}\right]$$

$$+\left(\tilde{u}+\frac{\gamma\sigma_\varepsilon^2}{x_I}\tilde{y}-\frac{Z\gamma\sigma_\varepsilon^2}{x_I}\right). \tag{2-8'}$$

Given the price $p_1$, we can derive the optimal demand of an informed trader is $\widetilde{\theta_I}=\frac{Z-\tilde{y}}{x_I}$ by Proposition 1. It is obvious that the market clearing condition holds if the demand of each informed and uninformed trader are $\widetilde{\theta_I}$ and $\widetilde{\theta_U}$ respectively. So, $\left(\widetilde{\theta_I},\widetilde{\theta_U},p_1\right)$ constitutes a market equilibrium in the case that investors acquire range information with asymmetry.

By equation (2-8’), it can be seen that the equilibrium price $p_1$ can be impacted by the range information acquired by informed traders, that is, $p_1$ depends on the upper bound $\overline{v}$ and the lower bound $\underline{v}$. However, as long as the range information $\left[\underline{v}_0,\ \overline{v}_0\right]$ acquired by uninformed traders is rougher than that acquired by informed traders, $\left[\underline{v}_0,\ \overline{v}_0\right]$ cannot affect the price.

## 3. The baseline model

We consider the market equilibrium under the case that investors cannot acquire any range information as a baseline, which has been discussed sufficiently by existing literature.

All the investors cannot obtain any range information, but informed traders can still receive the private signal $\tilde{u}$. Then the decision problem faced by a representative informed trader at t=0 is

$$\max_{\theta_I} E\left[-e^{-\gamma(D_0+\theta_I(v-p))} | u = \tilde{u}\right].$$

Its solution is

$$\theta_I = \frac{\tilde{u} - p}{\gamma \sigma_\varepsilon^2} \qquad (3-1)$$

referring to Grossman and Stigliz(1980), we conjecture the equilibrium price as a linear form

$$p = \tau\tilde{u} + \alpha\tilde{y} + \beta$$

Inserting it into equation (3-1) can derive

$$\theta_I = \frac{(1-\tau)\tilde{u} - \alpha\tilde{y} - \beta}{\gamma \sigma_\varepsilon^2} \qquad (3-2)$$

Based on the price $p$, uninformed traders can infer that

$$u + \frac{\alpha}{\tau} y = \frac{p - \beta}{\tau}$$

Thus, for a representative uninformed trader, the decision problem is

$$\max_{\theta_U} E\left[-e^{-\gamma(D_0+\theta_U(v-p))} \middle| u + \frac{\alpha}{\tau} y = \frac{p-\beta}{\tau}\right]$$

$$= \max_{\theta_U} -e^{-\gamma D_0 - \gamma(\mu_\eta - p)\theta_U + \frac{\gamma^2 \sigma_\eta^2}{2} \cdot \theta_U^2}$$

where $\mu_\eta = \omega_1\mu_0 + \omega_2\left(\frac{p-\beta}{\tau}\right), \sigma_\eta^2 = \sigma_\varepsilon^2 + \omega_1\sigma_u^2$, $\omega_1 = \frac{\alpha^2\sigma_y^2}{\tau^2\sigma_u^2+\alpha^2\sigma_y^2}, \omega_2 = 1 - \omega_1 = \frac{\tau^2\sigma_u^2}{\tau^2\sigma_u^2+\alpha^2\sigma_y^2}$ . The solution is

$$\theta_U = \frac{\mu_\eta - p}{\gamma\sigma_\eta^2}$$

$$= \frac{1}{\gamma\sigma_\eta^2}\left[\omega_1\mu_0 + (\omega_2 - \tau)\tilde{u} + \left(\frac{\omega_2}{\tau} - 1\right)\alpha\tilde{y} - \beta\right]. \qquad (3-3)$$

Substituting equation(3-2) and (3-3) into equation (2-1) and comparing the coefficient of $\tilde{u}$, $\tilde{y}$ and the intercept on the left-hand side with that on the right-hand side, we can derive

$$\tau = 1 - \frac{x_U}{1 + \frac{x_I^2\sigma_u^2}{\gamma^2\sigma_\varepsilon^4\sigma_y^2} + \frac{x_I\sigma_u^2}{\sigma_\varepsilon^2}}, \quad \alpha = \frac{\gamma\sigma_\varepsilon^2}{x_I}\cdot\tau, \quad \beta = \left(\frac{x_U\mu_0\omega_1}{\gamma\sigma_\eta^2} - Z\right)\Big/\left(\frac{x_U}{\gamma\sigma_\eta^2} + \frac{x_I}{\gamma\sigma_\varepsilon^2}\right).$$

In summary, without any release of range information, the equilibrium price is

$$p_0=\left(1-\frac{x_U}{1+\frac{x_I^2\sigma_u^2}{\gamma^2\sigma_\varepsilon^4\sigma_y^2}+\frac{x_I\sigma_u^2}{\sigma_\varepsilon^2}}\right)\tilde{u}+\frac{\gamma\sigma_\varepsilon^2}{x_I}\cdot\left(1-\frac{x_U}{1+\frac{x_I^2\sigma_u^2}{\gamma^2\sigma_\varepsilon^4\sigma_y^2}+\frac{x_I\sigma_u^2}{\sigma_\varepsilon^2}}\right)\tilde{y}$$

$$+\frac{\frac{x_U\mu_0\omega_1}{\gamma\sigma_\eta^2}-Z}{\frac{x_U}{\gamma\sigma_\eta^2}+\frac{x_I}{\gamma\sigma_\varepsilon^2}}. \qquad (3-4)$$

Under the equilibrium, the trading volume of informed and uninformed traders are all not zero.

## 4. The comparison between the two equilibriums

4.1 Factors affecting market price

The baseline model is developed under the assumption that investors cannot acquire any range information, whereas our main model is developed under the setup that investors acquire range information with asymmetry. Therefore, the comparison between the equilibriums of the two models can reveal the impact of investors' asymmetrical acquisition of range information on the risky asset market. We explain the intuition behind this point in the following.

In the baseline model, the equilibrium price $p_0$ (i.e., equation (3-4)) is affected by the precision of the private signal $\sigma_\varepsilon^2$, the informativeness of the private signal $\sigma_u^2$ and the size of noise trading $\sigma_y^2$. However, in our main model, the equilibrium price $p_1$ (i.e., equation (2-8')) is only affected by $\sigma_\varepsilon^2$, while $\sigma_u^2$ and $\sigma_y^2$ cannot have any impact on $p_1$. This implies that investors' asymmetrical acquisition of range information causes that the informativeness of the private signal $\sigma_u^2$ and the size of noise trading $\sigma_y^2$ can no longer affect the market price.

Since the realized value of $\varepsilon$ is unknown to all investors, their demands for the asset depend on its variance $\sigma_\varepsilon^2$ and then $\sigma_\varepsilon^2$ can affect the price.

Since the realized value of the private signal $\tilde{u}$ is known to informed traders, equation (3-4) and (2-8') imply that they can infer the realized value of the noise trading volume $\tilde{y}$ from the price. Thus, their demands do not depend on the variance of $u$ and $y$.

$\tilde{u}$ is unknown to uninformed traders, which prevents them from inferring $\tilde{y}$ from the price. As a result, their demands depend on the variance of $u$ and $y$.

In the baseline model, uninformed traders choose to participate in trading, which causes the price $p_0$ is affected by the variance of $u$ and $y$. In our main model, because the demands of informed traders do not depend on the variance of $u$ and $y$ and Proposition 2 shows that uninformed traders choose not to participate in trading, the market price $p_1$ will no longer be affected by $\sigma_u^2$ and $\sigma_y^2$.

## 4.2 The sensitivity of price to private signal

In the remaining, we suppose there always exists informed traders in the market, that is, $x_I > 0$ and then $0 \leq x_U < 1$.

We analyze the reaction of the market price when the private signal varies. In the baseline model, the sensitivity of the price to the private signal is

$$
\begin{aligned}
React_0 &\triangleq \frac{\partial p_0}{\partial \tilde{u}} \\
&= 1 - \frac{x_U}{1 + \frac{x_I^2 \sigma_u^2}{\gamma^2 \sigma_\varepsilon^4 \sigma_y^2} + \frac{x_I \sigma_u^2}{\sigma_\varepsilon^2}} > 0 \qquad (4-1)
\end{aligned}
$$

In our main model, by the item (2) in appendix A.2, the sensitivity is

$$
\begin{aligned}
React_1 &\triangleq \frac{\partial p_1}{\partial \tilde{u}} \\
&= \boldsymbol{H}_{[\underline{v},\overline{v}]}\left(\tilde{u} + \frac{\gamma \sigma_\varepsilon^2}{x_I}\tilde{y} - \frac{Z\gamma\sigma_\varepsilon^2}{x_I}\right) \\
&> 0, \qquad (4-2)
\end{aligned}
$$

where

$$
\boldsymbol{H}_{[\underline{v},\overline{v}]}(t) \triangleq \frac{1}{\sigma_\varepsilon^2}\left[\frac{\int_{\underline{v}-t}^{\overline{v}-t} x^2 f_\varepsilon(x)\,dx}{\int_{\underline{v}-t}^{\overline{v}-t} f_\varepsilon(x)\,dx} - \left(\frac{\int_{\underline{v}-t}^{\overline{v}-t} x f_\varepsilon(x)\,dx}{\int_{\underline{v}-t}^{\overline{v}-t} f_\varepsilon(x)\,dx}\right)^2\right].
$$

$React_0$ is a positive constant and $React_0 < 1$, which means it will not vary with the signal $\tilde{u}$ and noise trading volume $\tilde{y}$, and the variation in the price will be less than that in the signal when the signal varies. In the models of lots of related literature, the sensitivity of price to signals is also constant and positive, such as Kodres and Pritsker(2022), Easley et.al.(2014), Mondria et.al.(2022).

By equation (4-2) and the item (2) and (3) in appendix A.7, $React_1$ is also positive and $React_1 < 1$. However, according to the item (1) in appendix A.7, $React_1$ is not constant and varies continuously with the private signal $\tilde{u}$ and noise trading volume $\tilde{y}$. Figure.1 illustrates the variability and continuity of $React_1$.

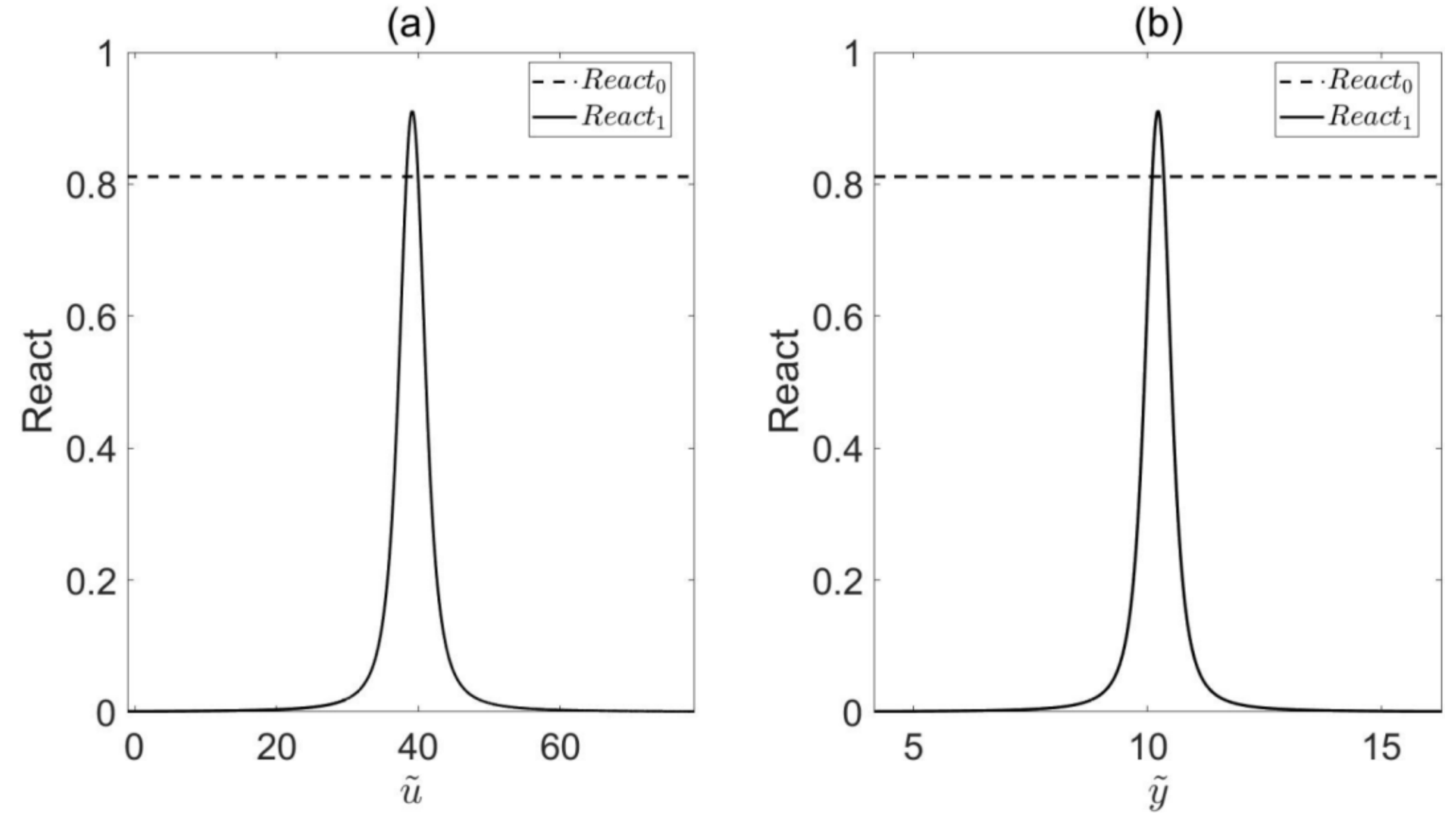

Figure.1. Take the value of $\gamma = 2, \sigma_\varepsilon^2 = 1, \sigma_u^2 = 9, \sigma_y^2 = 16, x_I = 0.3, Z = 10, \overline{v} = 15, \underline{v} = 10$. In panel (a), we take $\tilde{y} = 6$ and show the impact of increasing $\tilde{u}$ on the sensitivity of the price. In panel (b), we take $\tilde{u} = 11$ and show the impact of increasing $\tilde{y}$ on the sensitivity of the price.

In summary, the comparison implies that investors' acquisition of range information causes that the sensitivity of market price will vary continuously with private signal and noise trading volume and be no longer constant. In the models of related literature, the sensitivity of price to private signal is usually constant or piecewise constant. So our main model reveals a new pattern of market price varying with private signal, which is rarely recorded by existing literature.

In the case that investors don not acquire any range information, the sensitivity of the price $React_0$ depends on $\sigma_y^2, \sigma_u^2$ and $\sigma_\varepsilon^2$. An increase in $\sigma_\varepsilon^2$ or $\sigma_y^2$, or an decrease in $\sigma_u^2$ will lead to a decrease in $React_0$. In the case that investors acquire range information with asymmetry, the sensitivity $React_1$ still depends on $\sigma_\varepsilon^2$ but is independent of the informativeness of the private signal $\sigma_u^2$ and the size of noise trading $\sigma_y^2$. The reason is that uninformed traders whose demands depend on $\sigma_y^2$ and $\sigma_u^2$ do not trade assets in this case.

Remind that $React_1 = \boldsymbol{H}_{[\underline{v},\overline{v}]}\left(\tilde{u} + \frac{\gamma\sigma_\varepsilon^2}{x_I}\tilde{y} - \frac{Z\gamma\sigma_\varepsilon^2}{x_I}\right)$. According to the item (3) in appendix A.7, $0 < \boldsymbol{H}_{[\underline{v},\overline{v}]}(t) < 1$ holds for any $t \in R$. Use $S(\sigma_\varepsilon^2, \underline{v}, \overline{v})$ to denote the supremum of $\boldsymbol{H}_{[\underline{v},\overline{v}]}(\cdot)$, that is,

$$S(\sigma_\varepsilon^2,\underline{v},\overline{v}) \triangleq \underset{t\in R}{\text{Sup}}\, \boldsymbol{H}_{[\underline{v},\overline{v}]}(t).$$

Since $\boldsymbol{React_1}$ is not constant, $S(\sigma_\varepsilon^2,\underline{v},\overline{v})$ should be regarded as the supremum of the sensitivity of the price when investors asymmetrically acquire range information, which is only determined by three parameters: $\sigma_\varepsilon^2,\underline{v},\overline{v}$.

By equation (4-1), we have

$$0 < x_I = 1 - x_U < React_0 \le 1$$

Keep the supply $Z$ and the range information $[\underline{v},\overline{v}]$ fixed. According to the item (4) in appendix A.7, we have

$$\lim_{\left|\tilde{u}+\frac{\gamma\sigma_\varepsilon^2}{x_I}\tilde{y}\right|\to+\infty} \boldsymbol{H}_{[\underline{v},\overline{v}]}\left(\tilde{u}+\frac{\gamma\sigma_\varepsilon^2}{x_I}\tilde{y}-\frac{Z\gamma\sigma_\varepsilon^2}{x_I}\right)=0,$$

which implies $\boldsymbol{React_1} < \boldsymbol{React_0}$ holds if $\left|\tilde{u}+\frac{\gamma\sigma_\varepsilon^2}{x_I}\tilde{y}\right|$ is large enough. This manifests the asymmetrical acquisition of range information can decrease the sensitivity of price to private signal when the value of $\left|\tilde{u}+\frac{\gamma\sigma_\varepsilon^2}{x_I}\tilde{y}\right|$ is large enough.

By equation (4-1), if $1-\frac{x_U}{1+\frac{x_I^2\sigma_u^2}{\gamma^2\sigma_\varepsilon^4\sigma_y^2}+\frac{x_I\sigma_u^2}{\sigma_\varepsilon^2}} > S(\sigma_\varepsilon^2,\underline{v},\overline{v})$, that is, $\frac{x_I^2\sigma_u^2}{\gamma^2\sigma_\varepsilon^4\sigma_y^2}+\frac{x_I\sigma_u^2}{\sigma_\varepsilon^2} > \frac{S(\sigma_\varepsilon^2,\underline{v},\overline{v})-x_I}{1-S(\sigma_\varepsilon^2,\underline{v},\overline{v})}$, then $\boldsymbol{React_1}$ will be lower than $\boldsymbol{React_0}$ for all $\tilde{\boldsymbol{u}} \in \boldsymbol{R}$ and $\tilde{\boldsymbol{y}} \in \boldsymbol{R}$. If $1-\frac{x_U}{1+\frac{x_I^2\sigma_u^2}{\gamma^2\sigma_\varepsilon^4\sigma_y^2}+\frac{x_I\sigma_u^2}{\sigma_\varepsilon^2}} < S(\sigma_\varepsilon^2,\underline{v},\overline{v})$, that is, $\frac{x_I^2\sigma_u^2}{\gamma^2\sigma_\varepsilon^4\sigma_y^2}+\frac{x_I\sigma_u^2}{\sigma_\varepsilon^2} < \frac{S(\sigma_\varepsilon^2,\underline{v},\overline{v})-x_I}{1-S(\sigma_\varepsilon^2,\underline{v},\overline{v})}$, then $\boldsymbol{React_1}$ will exceed $\boldsymbol{React_0}$ for some values of private signal $\tilde{\boldsymbol{u}}$ and noise trading volume $\tilde{\boldsymbol{y}}$. The panel (a) and (b) of Figure.1 illustrate the case that $\boldsymbol{React_1}$ exceeds $\boldsymbol{React_0}$ in a small interval of $\tilde{\boldsymbol{u}}$ and $\tilde{\boldsymbol{y}}$ respectively.

In summary ,we propose the following proposition.

**Proposition 3.** If $1-\frac{x_U}{1+\frac{x_I^2\sigma_u^2}{\gamma^2\sigma_\varepsilon^4\sigma_y^2}+\frac{x_I\sigma_u^2}{\sigma_\varepsilon^2}} > S(\sigma_\varepsilon^2,\underline{v},\overline{v})$ , investors' asymmetrical acquisition of range information will definitely decrease the sensitivity of price to private signal. If $1-\frac{x_U}{1+\frac{x_I^2\sigma_u^2}{\gamma^2\sigma_\varepsilon^4\sigma_y^2}+\frac{x_I\sigma_u^2}{\sigma_\varepsilon^2}} < S(\sigma_\varepsilon^2,\underline{v},\overline{v})$, the acquisition can increase the sensitivity of price at some values of private signal $\tilde{u}$ and noise trading volume $\tilde{y}$ and decrease it if $\left|\tilde{u}+\frac{\gamma\sigma_\varepsilon^2}{x_I}\tilde{y}\right|$ reaches a large enough value.

For making the two conditions mentioned in Proposition 3 more concrete, we need to utilize the following lemma. Its proof is given in appendix A.8.

**Lemma 3.** For any $\underline{v}$, $\overline{v}$ with $-\infty<\underline{v}<\overline{v}<+\infty$, $0<\mathrm{S}(\sigma_\varepsilon^2,\underline{v},\overline{v})<1$ holds.

By Lemma 3, we have $\mathrm{S}(\sigma_\varepsilon^2,\underline{v},\overline{v})<1$ and remind that

$$x_I<React_0\le 1.$$

If the proportion of informed traders $x_I$ is large such that $x_I>\mathrm{S}(\sigma_\varepsilon^2,\underline{v},\overline{v})$, it follows that $React_0>\mathrm{S}(\sigma_\varepsilon^2,\underline{v},\overline{v})$, which means $React_0>React_1$ holds for all $\tilde{u}\in\boldsymbol{R}$ and $\tilde{y}\in\boldsymbol{R}$.

Consider the situation that $x_I<\mathrm{S}(\sigma_\varepsilon^2,\underline{v},\overline{v})$. By equation (4-1), we have

$$\lim_{\sigma_u^2\to 0} React_0=x_I.$$

which implies that if the informativeness of the private signal $\sigma_u^2$ is small enough, $React_0<\mathrm{S}(\sigma_\varepsilon^2,\underline{v},\overline{v})$ can hold, that is, $React_1>React_0$ holds for some values of $\tilde{u}$ and $\tilde{y}$.

Since $x_I+x_U=1$, $x_I$ approaches 0 when $x_U$ approaches 1. By equation (4-1), we can derive

$$\lim_{x_U\to 1} React_0=0.$$

Remind that lemma 3 indicates $\mathrm{S}(\sigma_\varepsilon^2,\underline{v},\overline{v})>0$. Thus, if the proportion of informed traders $x_I$ is small enough, $React_0<\mathrm{S}(\sigma_\varepsilon^2,\underline{v},\overline{v})$ can hold, that is, $React_1>React_0$ holds for some values of $\tilde{u}$ and $\tilde{y}$.

By equation (4-1), we can derive $\lim_{\sigma_\varepsilon^2\to+\infty} React_0=x_I$. By the item (5) in appendix A.7,

$$React_1=\boldsymbol{H}_{[\underline{v},\overline{v}]}\left(\tilde{u}+\frac{\gamma\sigma_\varepsilon^2}{x_I}\tilde{y}-\frac{Z\gamma\sigma_\varepsilon^2}{x_I}\right)\le\frac{(\overline{v}-\underline{v})^2}{\sigma_\varepsilon^2}$$

holds for all $\tilde{u}\in\boldsymbol{R}$ and $\tilde{y}\in\boldsymbol{R}$. Given $\lim_{\sigma_\varepsilon^2\to+\infty}\frac{(\overline{v}-\underline{v})^2}{\sigma_\varepsilon^2}=0$, $React_1<React_0$

holds for all $\tilde{u} \in \boldsymbol{R}$ and $\tilde{y} \in \boldsymbol{R}$ if the deviation of the private signal $\sigma_\varepsilon^2$ is large enough, that is, the precision of the signal is low enough.

By equation (4-1), we have

$$\lim_{\sigma_u^2 \to +\infty} React_0 = 1 \text{ and } \lim_{\sigma_y^2 \to 0} React_0 = 1.$$

Since $\mathrm{S}(\sigma_\varepsilon^2, \underline{v}, \overline{v}) < 1$, we have $React_0 > \mathrm{S}(\sigma_\varepsilon^2, \underline{v}, \overline{v})$ if the informativeness of the private signal $\sigma_u^2$ is large enough and the size of noise trading $\sigma_y^2$ is small enough.

Consider the case that $x_I < 1$ (i.e., $x_U > 0$). It follows that $React_0 < 1$. By the item (3) in appendix A.7, we have

$$\begin{aligned}\lim_{\substack{\underline{v} \to -\infty \\ \overline{v} \to +\infty}} \mathrm{S}(\sigma_\varepsilon^2, \underline{v}, \overline{v}) &\geq \lim_{\substack{\underline{v} \to -\infty \\ \overline{v} \to +\infty}} React_1 \\ &= \lim_{\substack{\underline{v} \to -\infty \\ \overline{v} \to +\infty}} \boldsymbol{H}_{[\underline{v}, \overline{v}]}\left(\tilde{u} + \frac{\gamma \sigma_\varepsilon^2}{x_I} \tilde{y} - \frac{Z \gamma \sigma_\varepsilon^2}{x_I}\right) \\ &= 1 > React_0.\end{aligned}$$

Thus, if the range information acquired by informed traders is rough enough, that is, $\underline{v}$ is low enough and $\overline{v}$ is high enough, then $React_1 > React_0$ holds for some values of $\tilde{u}$ and $\tilde{y}$.

The analysis above reveals that investors' asymmetrical acquisition of range information can increase the sensitivity of price under some conditions and decrease them under some other conditions. In summary, we propose two corollaries in the following.

**Corollary 1.** If one of the following conditions holds, then investors' asymmetrical acquisition of range information will definitely decrease the sensitivity of market price to private signal.

(1) The proportion of informed traders is large such that $x_I > \mathrm{S}(\sigma_\varepsilon^2, \underline{v}, \overline{v})$.

(2) The precision of private signal is low enough.

(3) The informativeness of private signal is large enough.

(4) The size of noise trading is small enough.

**Corollary 2.** If one of the following conditions holds, then investors' asymmetrical acquisition of range information can increase the sensitivity of price at some values of private signal $\tilde{u}$ and noise trading volume $\tilde{y}$.

(1) The proportion of informed traders $x_I < \mathrm{S}(\sigma_\varepsilon^2, \underline{v}, \overline{v})$ and the informativeness of private signal $\sigma_u^2$ is small enough.

(2) The proportion of informed traders is small enough.

(3) The proportion of uninformed traders is not 0 and the range information acquired by informed traders is rough enough, that is, $\underline{v}$ is low enough and $\overline{v}$ is high enough.

Informed traders can directly receive the private signal, whereas uninformed traders can only infer it from the price and the outcome of their inference is polluted by noise trading. So uninformed traders' reactions to the private signal are smaller than informed traders'. Investors' asymmetrical acquisition of range information can generate two effects on the sensitivity of price and their directions are opposite.

First, by Proposition 2, uninformed traders who can only acquire rougher range information choose not to trade assets, while informed traders still keep trading. Because uninformed traders' reactions to the private signal are smaller, their exit from the trading can make the reaction of price to the signal increase, that is, the sensitivity of price rises. Second, once investors can acquire range information besides the private signal, the degree to which their decisions depend on the private signal will decline, which will reduce their reaction to the signal and then the sensitivity of price to the signal.

Under the four conditions given in Corollary 1, the first effect of the acquisition exceeds the second, and then the acquisition will increase the sensitivity of price. Under the three conditions given in Corollary 2, the second effect exceeds the first at some values of private signal and noise trading volume, and then the acquisition will decrease the sensitivity.

## 4.3 Market Liquidity

According to related literature, such as Goldstein and Yang(2017), market liquidity is defined as

$$Liquidity \triangleq \left(\frac{\partial p}{\partial \tilde{y}}\right)^{-1}.$$

A smaller Liquidity means noise trading $y$ has a smaller impact on market price and the market is regarded as deeper and more liquid.

In the equilibrium of the baseline model, the market liquidity is

$$Liquidity_0 = \left(\frac{\partial p_0}{\partial \tilde{y}}\right)^{-1}$$

$$= \frac{x_I}{\gamma\sigma_\varepsilon^2}\left(1 - \frac{x_U}{1 + \frac{x_I^2\sigma_u^2}{\gamma^2\sigma_\varepsilon^4\sigma_y^2} + \frac{x_I\sigma_u^2}{\sigma_\varepsilon^2}}\right)^{-1} > 0. \qquad (4-3)$$

In the equilibrium of our main model, by the item (2) in appendix A.2, the market liquidity is

$$Liquidity_1 = \left(\frac{\partial p_1}{\partial \tilde{y}}\right)^{-1}$$

$$= \frac{x_I}{\gamma\sigma_\varepsilon^2} \cdot \frac{1}{\boldsymbol{H}_{[\underline{v},\bar{v}]}\left(\tilde{u} + \frac{\gamma\sigma_\varepsilon^2}{x_I}\tilde{y} - \frac{Z\gamma\sigma_\varepsilon^2}{x_I}\right)}$$

$$= \frac{x_I}{\gamma\sigma_\varepsilon^2} \cdot \frac{1}{React_1} > 0 \qquad (4-4)$$

$Liquidity_0$ is constant, that is, it cannot vary with the private signal $\tilde{u}$ and noise trading $\tilde{y}$. Remind that $React_1$ is not constant and varies continuously with $\tilde{u}$ and $\tilde{y}$, and so does $Liquidity_1$ by equation (4-4). The comparison between $Liquidity_0$ and $Liquidity_1$ shows that investors' acquisition of range information can cause that market liquidity will vary continuously with private signal and noise trading volume and be no longer constant.

By equation (4-4), the range information acquired by uninformed traders, that is, the upper bound $\bar{v}_0$ and the lower bound $\underline{v}_0$, cannot have any impact on the market liquidity.

$Liquidity_0$ decreases with the informativeness of the private signal $\sigma_u^2$ and increases with the size of noise trading $\sigma_y^2$, whereas $Liquidity_1$ is independent of $\sigma_u^2$ and $\sigma_y^2$. Thus, investors' asymmetrical acquisition of range information causes that market liquidity can no longer be impacted by the informativeness of the private signal and the size of noise trading.

By equation (4-4), the infimum of $Liquidity_1$ is $\frac{x_I}{\gamma\sigma_\varepsilon^2} \cdot \frac{1}{S(\sigma_\varepsilon^2, \underline{v}, \bar{v})}$. The item (4) in appendix A.7 implies that

$$\lim_{\left|\tilde{u} + \frac{\gamma\sigma_\varepsilon^2}{x_I}\tilde{y}\right| \to +\infty} Liquidity_1 = +\infty.$$

This means that the market will become very liquid, that is, noise trading will have little impact on market price, when the value of $\left|\tilde{u} + \frac{\gamma\sigma_\varepsilon^2}{x_I}\tilde{y}\right|$ is large enough. In summary, we have

$$+\infty > Liquidity_1 \geq \frac{x_I}{\gamma\sigma_\varepsilon^2} \cdot \frac{1}{\mathrm{S}\left(\sigma_\varepsilon^2, \underline{v}, \overline{v}\right)}$$

We give the following proposition to clarify the impact of investors' acquisition of range information on market liquidity.

**Proposition 4.** If $1 - \frac{x_U}{1 + \frac{x_I^2\sigma_u^2}{\gamma^2\sigma_\varepsilon^4\sigma_y^2} + \frac{x_I\sigma_u^2}{\sigma_\varepsilon^2}} > \mathrm{S}\left(\sigma_\varepsilon^2, \underline{v}, \overline{v}\right)$, that is, $\frac{x_I^2\sigma_u^2}{\gamma^2\sigma_\varepsilon^4\sigma_y^2} + \frac{x_I\sigma_u^2}{\sigma_\varepsilon^2} > \frac{\mathrm{S}(\sigma_\varepsilon^2, \underline{v}, \overline{v}) - x_I}{1 - \mathrm{S}(\sigma_\varepsilon^2, \underline{v}, \overline{v})}$, investors' asymmetrical acquisition of range information will definitely increase the market liquidity. If $1 - \frac{x_U}{1 + \frac{x_I^2\sigma_u^2}{\gamma^2\sigma_\varepsilon^4\sigma_y^2} + \frac{x_I\sigma_u^2}{\sigma_\varepsilon^2}} < \mathrm{S}\left(\sigma_\varepsilon^2, \underline{v}, \overline{v}\right)$, that is, $\frac{x_I^2\sigma_u^2}{\gamma^2\sigma_\varepsilon^4\sigma_y^2} + \frac{x_I\sigma_u^2}{\sigma_\varepsilon^2} < \frac{\mathrm{S}(\sigma_\varepsilon^2, \underline{v}, \overline{v}) - x_I}{1 - \mathrm{S}(\sigma_\varepsilon^2, \underline{v}, \overline{v})}$, the acquisition can decrease the market liquidity at some values of private signal $\tilde{u}$ and noise trading volume $\tilde{y}$ and increase it if $\left|\tilde{u} + \frac{\gamma\sigma_\varepsilon^2}{x_I}\tilde{y}\right|$ reaches a large enough value.

For making the two conditions mentioned in Proposition 4 more concrete, we can derive the following two corollaries by an analysis similar to that in Section 4.2.

**Corollary 3.** If one of the following conditions holds, then investors' asymmetrical acquisition of range information will definitely increase market liquidity.

(1) The proportion of informed traders is large such that $x_I > \mathrm{S}\left(\sigma_\varepsilon^2, \underline{v}, \overline{v}\right)$.

(2) The precision of private signal is low enough.

(3) The informativeness of private signal is large enough.

(4) The size of noise trading is small enough.

**Corollary 4.** If one of the following conditions holds, then investors' asymmetrical acquisition of range information can decrease market liquidity at some values of private signal $\tilde{u}$ and noise trading volume $\tilde{y}$.

(1) The proportion of informed traders $x_I < \mathrm{S}\left(\sigma_\varepsilon^2, \underline{v}, \overline{v}\right)$ and the informativeness of private signal $\sigma_u^2$ is small enough.

(2) The proportion of informed traders is small enough.

(3) The proportion of uninformed traders is not 0 and the range information acquired by informed traders is rough enough.

The completeness of investors' information and the information asymmetry among investors are two important factors that can decide market liquidity. Generally speaking, market liquidity increases with the completeness of investors' information and decreases with the degree of information asymmetry.

Investors' asymmetrical acquisition of range information can generate two effects on market liquidity and their directions are opposite.

First, the acquisition of range information can improve the completeness of investors' information and then raise the market liquidity. Second, since the range information acquired by uninformed traders is rougher than that acquired by informed traders, the asymmetrical acquisition will add the degree of information asymmetry and then decrease market liquidity.

Under the four conditions given in Corollary 3, the first effect of the acquisition exceeds the second, and then the acquisition will increase market liquidity. Under the three conditions given in Corollary 4, the second effect exceeds the first at some values of private signal and noise trading volume, and then the acquisition will decrease market liquidity.

4.4 Market Efficiency

In related literature, market efficiency is usually measured with price informativeness. A larger price informativeness implies a higher market efficiency. Price informativeness is defined as $PI \triangleq \frac{1}{Var(v|p)}$.

For the baseline model, the price informativeness is

$$PI_0 = \frac{1}{Var(v|p_0)} = \left[Var\left(v \middle| u + \frac{\gamma\sigma_\varepsilon^2}{x_I}y\right)\right]^{-1} = \left[\sigma_\varepsilon^2 + \frac{\gamma^2\sigma_\varepsilon^4\sigma_y^2\sigma_u^2}{x_I^2\sigma_u^2 + \gamma^2\sigma_\varepsilon^4\sigma_y^2}\right]^{-1}.$$

For our main model, by equation (2-8), the price informativeness is

$$\begin{aligned} PI_1 &= \frac{1}{Var(v|p_1)} \\ &= \left[Var\left(v \middle| J_{[\underline{v},\ \bar{v}]}\left(u + \frac{\gamma\sigma_\varepsilon^2}{x_I}y - \frac{Z\gamma\sigma_\varepsilon^2}{x_I}\right)\right)\right]^{-1} \\ &= \left[Var\left(v \middle| u + \frac{\gamma\sigma_\varepsilon^2}{x_I}y\right)\right]^{-1} \\ &= \left[\sigma_\varepsilon^2 + \frac{\gamma^2\sigma_\varepsilon^4\sigma_y^2\sigma_u^2}{x_I^2\sigma_u^2 + \gamma^2\sigma_\varepsilon^4\sigma_y^2}\right]^{-1}. \end{aligned}$$

Thus, $PI_0 = PI_1$. This implies that investors' acquisition of range information cannot have any impact on market efficiency.

## 5. Conclusion

Our main model analyzes investors' trading behavior, market equilibrium, the sensitivity of price to private signal, market liquidity and efficiency under the case that they acquire range information with asymmetry.

Originally, uninformed traders don not have direct access to the private signal which is held by informed traders. Our study suggests that once range information is released to investors with asymmetry, uninformed traders who can only acquire rougher range information will give up trading under the max-min ambiguity aversion criterion. We reveal that investors' asymmetrical acquisition of range information can cause that market liquidity and the sensitivity of market price to private signal is no longer constant and vary continuously with private signal and noise trading volume, which is rarely recorded by existing literature.

As a benchmark, we use a baseline model which describes the equilibrium under the case that investors cannot acquire any range information. By comparing the equilibriums of the two models, we reveal that investors' asymmetrical acquisition of range information will cause that the informativeness of private signal and the size of noise trading cannot have any impact on market price and liquidity.

Additionally, our comparison also reveals that investors' asymmetrical acquisition of range information may increase the sensitivity of price and market liquidity or reduce them, which depends on three types of factors in the following.

(1)Some parameters of market, such as the informativeness of private signal, the size of noise trading, the proportion of informed traders.

(2)The value of private signal and noise trading volume.

(3)The range information, that is, the lower bound and upper bound.

We also propose the conditions where the acquisition can increase and reduce the sensitivity and market liquidity respectively.

As far as practition, this study indicates that if we release range information to market with asymmetry, the release will not necessarily reduce the sensitivity of price to private signal and raise market liquidity. If investors' acquisition of range information is inevitably asymmetrical, whether to release it should depend on the level of three type of factors listed above. If the range information acquired by uninformed traders is throughout rougher, improving the precision of their range information will not affect market price and quality.

## Appendix

A.1 The calculation of equation (2-2)

Define a conditional distribution $w \triangleq v|(u=\tilde{u})$, then $w \sim N(\tilde{u}, \sigma_\varepsilon^2)$. The distribution and density function of $w$ are denoted by $F_w$ and $f_w$ respectively. The denominator of equation (2-2) is

$$E\left[\mathbf{1}_{\{v\in[\underline{v},\overline{v}]\}}|u=\tilde{u}\right]$$

$$=E\left[\mathbf{1}_{\{w\in[\underline{v},\overline{v}]\}}\right]$$

$$=\int_{\underline{v}}^{\overline{v}}\frac{1}{\sqrt{2\pi}\sigma_\varepsilon}e^{-\frac{(x-\tilde{u})^2}{2\sigma_\varepsilon^2}}dx$$

$$=\Psi\left(\frac{\overline{v}-\tilde{u}}{\sigma_\varepsilon}\right)-\Psi\left(\frac{\underline{v}-\tilde{u}}{\sigma_\varepsilon}\right).$$

The denominator of equation (2-2) is

$$E\left[-e^{-\gamma(D_0+\theta_I(v-p))}\mathbf{1}_{\{v\in[\underline{v},\overline{v}]\}}|u=\tilde{u}\right]$$

$$=E\left[-e^{-\gamma(D_0+\theta_I(w-p))}\mathbf{1}_{\{w\in[\underline{v},\overline{v}]\}}\right]$$

$$=-e^{-\gamma D_0+\gamma p\theta_I}\int_{\underline{v}}^{\overline{v}}e^{-\gamma\theta_I x}\cdot\frac{1}{\sqrt{2\pi}\sigma_\varepsilon}e^{-\frac{(x-\tilde{u})^2}{2\sigma_\varepsilon^2}}dx$$

$$=-e^{-\gamma D_0+\gamma(p-\tilde{u})\theta_I+\frac{\gamma^2\sigma_\varepsilon^2\theta_I^2}{2}}\left[F_w(\overline{v}+\gamma\sigma_\varepsilon^2\theta_I)-F_w(\underline{v}+\gamma\sigma_\varepsilon^2\theta_I)\right]$$

$$=-e^{-\gamma D_0+\gamma(p-\tilde{u})\theta_I+\frac{\gamma^2\sigma_\varepsilon^2\theta_I^2}{2}}\left[\Psi\left(\frac{\overline{v}+\gamma\sigma_\varepsilon^2\theta_I-\tilde{u}}{\sigma_\varepsilon}\right)-\Psi\left(\frac{\underline{v}+\gamma\sigma_\varepsilon^2\theta_I-\tilde{u}}{\sigma_\varepsilon}\right)\right].$$

A.2 Some properties of the function $\boldsymbol{J}_{[\underline{v},\ \overline{v}]}(t)$

Remind that $\varepsilon \sim N(0,\sigma_\varepsilon^2)$. Its distribution and density function are denoted by $F_\varepsilon$ and $f_\varepsilon$ respectively. For any fixed $\underline{v}$, $\overline{v}\in R$ with $\overline{v}>\underline{v}$, the following properties hold.

(1) $\underline{v}<\boldsymbol{J}_{[\underline{v},\ \overline{v}]}(t)<\overline{v}$.

(2) $\boldsymbol{J}_{[\underline{v},\ \overline{v}]}'(t)=\boldsymbol{H}_{[\underline{v},\overline{v}]}(t)$, where

$$\boldsymbol{H}_{[\underline{v},\overline{v}]}(t)\triangleq\frac{1}{\sigma_\varepsilon^2}\left[\frac{\int_{\underline{v}-t}^{\overline{v}-t}x^2f_\varepsilon(x)\,dx}{\int_{\underline{v}-t}^{\overline{v}-t}f_\varepsilon(x)\,dx}-\left(\frac{\int_{\underline{v}-t}^{\overline{v}-t}xf_\varepsilon(x)\,dx}{\int_{\underline{v}-t}^{\overline{v}-t}f_\varepsilon(x)\,dx}\right)^2\right].$$

(3) $1>\boldsymbol{J}_{[\underline{v},\ \overline{v}]}'(t)>0$.

(4) $\lim_{t\to+\infty} \boldsymbol{J}_{[\underline{v},\ \overline{v}]}(t) = \overline{v}$ and $\lim_{t\to-\infty} \boldsymbol{J}_{[\underline{v},\ \overline{v}]}(t) = \underline{v}$.

(5) $\boldsymbol{J}_{[\underline{v},\ \overline{v}]}(t)$ is continuous with respect to $t, \underline{v}, \overline{v}$.

A.3 The proof of Proposition 1

The optimal demand $\widetilde{\theta_I}$ satisfies equation (2-3)

$$\boldsymbol{J}_{[\underline{v},\ \overline{v}]}\left(\tilde{u} - \gamma\sigma_\varepsilon^2\widetilde{\theta_I}\right) - p = 0,$$

which means $\widetilde{\theta_I}$ depends on $\tilde{u}$. For a given price $p$, by the implicit function theorem, we have

$$\frac{d\widetilde{\theta_I}}{d\tilde{u}} = -\frac{\dfrac{\partial \boldsymbol{J}_{[\underline{v},\ \overline{v}]}\left(\tilde{u} - \gamma\sigma_\varepsilon^2\widetilde{\theta_I}\right) - p}{\partial \tilde{u}}}{\dfrac{\partial \boldsymbol{J}_{[\underline{v},\ \overline{v}]}\left(\tilde{u} - \gamma\sigma_\varepsilon^2\widetilde{\theta_I}\right) - p}{\partial \widetilde{\theta_I}}}$$

$$= -\frac{\boldsymbol{J}_{[\underline{v},\ \overline{v}]}'\left(\tilde{u} - \gamma\sigma_\varepsilon^2\widetilde{\theta_I}\right)}{\boldsymbol{J}_{[\underline{v},\ \overline{v}]}'\left(\tilde{u} - \gamma\sigma_\varepsilon^2\widetilde{\theta_I}\right) \cdot (-\gamma\sigma_\varepsilon^2)}$$

$$= \frac{1}{\gamma\sigma_\varepsilon^2}.$$

So, we derive a differential equation

$$\frac{d\widetilde{\theta_I}}{d\tilde{u}} = \frac{1}{\gamma\sigma_\varepsilon^2},$$

the solution of which is

$$\widetilde{\theta_I} = \frac{1}{\gamma\sigma_\varepsilon^2} \cdot \tilde{u} + k,$$

where the constant $k$ is independent of $\theta_I$ and $\tilde{u}$. Substituting the solution of $\widetilde{\theta_I}$ into equation (2-3), we can derive

$$p = \boldsymbol{J}_{[\underline{v},\ \overline{v}]}(-\gamma\sigma_\varepsilon^2 k),$$

that is, equation (2-4).

For any fixed $p \in (\underline{v}, \overline{v})$, it can be proved that there exists a unique $k$ satisfying equation (2-4). By the item (3) in appendix A.2,

$$\frac{\partial \boldsymbol{J}_{[\underline{v},\ \overline{v}]}(-\gamma\sigma_\varepsilon^2 k)}{\partial k}$$

$$= -\gamma\sigma_\varepsilon^2 \cdot \boldsymbol{J}_{[\underline{v},\ \overline{v}]}'(-\gamma\sigma_\varepsilon^2 k)$$

$$< 0.$$

which implies that $\boldsymbol{J}_{[\underline{v},\ \overline{v}]}(-\gamma\sigma_\varepsilon^2 k)$ strictly decreases with $k$. By item (4) in appendix A.2,

$$\lim_{k\to+\infty}\boldsymbol{J}_{[\underline{v},\ \overline{v}]}(-\gamma\sigma_\varepsilon^2 k)=\underline{v}, \qquad \lim_{k\to-\infty}\boldsymbol{J}_{[\underline{v},\ \overline{v}]}(-\gamma\sigma_\varepsilon^2 k)=\overline{v}.$$

Therefore, there exists a unique $k$ such that $p=\boldsymbol{J}_{[\underline{v},\ \overline{v}]}(-\gamma\sigma_\varepsilon^2 k)$.

By the implicit function theorem,

$$\frac{dk}{dp}=\left[\frac{\partial\boldsymbol{J}_{[\underline{v},\ \overline{v}]}(-\gamma\sigma_\varepsilon^2 k)}{\partial k}\right]^{-1}<0$$

which implies that $\widetilde{\theta}_I=\frac{1}{\gamma\sigma_\varepsilon^2}\cdot\tilde{u}+k$ strictly decreases with $p$.

A.4 The calculation of equation (2-7)

Define a conditional distribution $\eta\triangleq v\left|\left(u+\frac{\alpha}{\tau}y=\frac{\boldsymbol{J}_{[\underline{v'},\ \overline{v'}]}^{-1}(p)-\beta}{\tau}\right)\right.$. It follows that $\eta\sim N(\mu_\eta,\sigma_\eta^2)$.

$$U_U\left(\theta_U;\underline{v}',\overline{v}'\right)$$

$$=E\left[-e^{-\gamma(D_0+\theta_U(v-p))}\left|u+\frac{\alpha}{\tau}y=\frac{\boldsymbol{J}_{[\underline{v'},\overline{v'}]}^{-1}(p)-\beta}{\tau},v\in\left[\underline{v}',\overline{v}'\right]\right.\right]$$

$$=\frac{E\left[-e^{-\gamma(D_0+\theta_U(v-p))}\mathbf{1}_{\{v\in[\underline{v}',\overline{v'}]\}}\left|u+\frac{\alpha}{\tau}y=\frac{\boldsymbol{J}_{[\underline{v'},\ \overline{v'}]}^{-1}(p)-\beta}{\tau}\right.\right]}{E\left[\mathbf{1}_{\{v\in[\underline{v}',\overline{v'}]\}}\left|u+\frac{\alpha}{\tau}y=\frac{\boldsymbol{J}_{[\underline{v'},\ \overline{v'}]}^{-1}(p)-\beta}{\tau}\right.\right]}$$

$$=\frac{E\left[-e^{-\gamma(D_0+\theta_U(\eta-p))}\mathbf{1}_{\{\eta\in[\underline{v}',\overline{v'}]\}}\right]}{E\left[\mathbf{1}_{\{\eta\in[\underline{v}',\overline{v'}]\}}\right]}$$

$$=\frac{-e^{-\gamma D_0+\gamma p\theta_U}\cdot\int_{\underline{v'}}^{\overline{v'}}e^{-\gamma\theta_U x}\cdot\frac{1}{\sqrt{2\pi}\sigma_\eta}e^{-\frac{(x-\mu_\eta)^2}{2\sigma_\eta^2}}dx}{\Psi\left(\frac{\overline{v}'-\mu_\eta}{\sigma_\eta}\right)-\Psi\left(\frac{\underline{v}'-\mu_\eta}{\sigma_\eta}\right)}$$

$$=\frac{-e^{-\gamma D_0+\gamma(p-\mu_\eta)\theta_U+\frac{\gamma^2\sigma_\eta^2\theta_U^2}{2}}\cdot\left[\Psi\left(\frac{\overline{v}'+\gamma\sigma_\eta^2\theta_U-\mu_\eta}{\sigma_\eta}\right)-\Psi\left(\frac{\underline{v}'+\gamma\sigma_\eta^2\theta_U-\mu_\eta}{\sigma_\eta}\right)\right]}{\Psi\left(\frac{\overline{v}'-\mu_\eta}{\sigma_\eta}\right)-\Psi\left(\frac{\underline{v}'-\mu_\eta}{\sigma_\eta}\right)}$$

A.5 The proof of Lemma 1

$$\frac{d\,U_U(\theta_U;\underline{v}',\overline{v}')}{d\theta_U}$$

$$= -\gamma\cdot U_U(\theta_U;\underline{v}',\overline{v}')\cdot\left[\boldsymbol{J}_{[\underline{v}',\ \overline{v}']}(\mu_\eta-\gamma\sigma_\eta^2\theta_U)-p\right].$$

Define $F(\theta_U)\triangleq\boldsymbol{J}_{[\underline{v}',\ \overline{v}']}(\mu_\eta-\gamma\sigma_\eta^2\theta_U)-p$ and $\bar{\theta}_U\triangleq\frac{1}{\gamma\sigma_\eta^2}\left[\omega_1\mu_0-\frac{\omega_2\beta}{\tau}+\left(\frac{\omega_2}{\tau}-1\right)\cdot\boldsymbol{J}^{-1}_{[\underline{v}',\overline{v}']}(p)\right]$. By the definition of $\boldsymbol{J}^{-1}_{[\underline{v}',\ \overline{v}']}(\cdot)$, it can be verified that $F(\bar{\theta}_U)=0$. By the item (3) in appendix A.2, we derive

$$F'(\theta_U)=-\gamma\sigma_\eta^2\cdot\boldsymbol{J}_{[\underline{v}',\ \overline{v}']}{}'(\mu_\eta-\gamma\sigma_\eta^2\theta_U)<0,$$

which means $F(\theta_U)$ is a strictly decreasing function of $\theta_U$.

So, for any $\theta_U<\bar{\theta}_U$, we have

$$\frac{d\,U_U(\theta_U;\underline{v}',\overline{v}')}{d\theta_U}=-\gamma\cdot U_U(\theta_U;\underline{v}',\overline{v}')\cdot F(\theta_U)>0.$$

For any $\theta_U>\bar{\theta}_U$, we have

$$\frac{d\,U_U(\theta_U;\underline{v}',\overline{v}')}{d\theta_U}=-\gamma\cdot U_U(\theta_U;\underline{v}',\overline{v}')\cdot F(\theta_U)<0.$$

A.6 The proof of Lemma 2

For a fixed $\underline{v}'\in(\underline{v}_0,p)$, hypothesize that $\lim_{\delta\to0+}\boldsymbol{J}^{-1}_{[\underline{v}',\ p+\delta]}(p)=+\infty$ does not hold. Under this hypothesis, there exists a positive constant $M$ and a positive decreasing sequence $\{\delta_n\}_{n=1}^{+\infty}$ satisfying $\lim_{n\to+\infty}\delta_n=0$ such that

$$\boldsymbol{J}^{-1}_{[\underline{v}',\ p+\delta_n]}(p)<M \qquad (A-1)$$

holds for any $n\in\boldsymbol{N}^+$.

By the item (3) in appendix A.2, $\boldsymbol{J}_{[\underline{v}',\ p+\delta_n]}{}'(t)>0$, which means $\boldsymbol{J}_{[\underline{v}',\ p+\delta_n]}(\cdot)$ is a strictly increasing function. According to $(A-1)$, we derive

$$p=\boldsymbol{J}_{[\underline{v}',\ p+\delta_n]}\left(\boldsymbol{J}^{-1}_{[\underline{v}',\ p+\delta_n]}(p)\right)<\boldsymbol{J}_{[\underline{v}',\ p+\delta_n]}(M).$$

By the item (5) in appendix A.2, $\boldsymbol{J}_{[\underline{v}',\ \overline{v}']}(t)$ is also continuous with respect to $\underline{v}'$ and $\overline{v}'$. Thus, let $n\to+\infty$, we have

$$p \le \boldsymbol{J}_{[\underline{v}',\ p]}(M). \qquad (A-2)$$

On the other hand, by the item (1) in appendix A.2,

$$\boldsymbol{J}_{[\underline{v}',\ p]}(M) < p. \qquad (A-3)$$

However, $(A-2)$ is contradictory with $(A-3)$, which implies the hypothesis does not hold and then $\lim_{\delta\to 0+} \boldsymbol{J}^{-1}_{[\underline{v}',\ p+\delta]}(p) = +\infty$.

In a similar way, it can be proved that $\lim_{\delta\to 0+} \boldsymbol{J}^{-1}_{[p-\delta,\ \overline{v}']}(p) = -\infty$ holds for any fixed $\overline{v}' \in (p, \overline{v}_0)$.

A.7 Some properties of the function $\boldsymbol{H}_{[\underline{v},\overline{v}]}(t)$

For any fixed $\underline{v}$, $\overline{v} \in R$ with $\overline{v} > \underline{v}$, the following properties hold.

(1) $\boldsymbol{H}_{[\underline{v},\overline{v}]}(t)$ is continuous with respect to $t$, $\underline{v}$, $\overline{v}$.

(2) For any $t \in R$, $\boldsymbol{H}_{[\underline{v},\overline{v}]}(t) > 0$.

(3) $\boldsymbol{H}_{[\underline{v},\overline{v}]}(t) < 1$ and $\lim_{\substack{\underline{v}\to-\infty \\ \overline{v}\to+\infty}} \boldsymbol{H}_{[\underline{v},\overline{v}]}(t) = 1$.

(4) $\lim_{t\to+\infty} \boldsymbol{H}_{[\underline{v},\overline{v}]}(t) = 0$ and $\lim_{t\to-\infty} \boldsymbol{H}_{[\underline{v},\overline{v}]}(t) = 0$.

(5) For any $t \in R$, $\boldsymbol{H}_{[\underline{v},\overline{v}]}(t) \le \frac{(\overline{v}-\underline{v})^2}{\sigma_\varepsilon^2}$.

The proof of the item (5).

$$\boldsymbol{H}_{[\underline{v},\overline{v}]}(t) = \frac{1}{\sigma_\varepsilon^2}\left[\frac{\int_{\underline{v}-t}^{\overline{v}-t} x^2 f_\varepsilon(x)\,dx}{\int_{\underline{v}-t}^{\overline{v}-t} f_\varepsilon(x)\,dx} - \left(\frac{\int_{\underline{v}-t}^{\overline{v}-t} x f_\varepsilon(x)\,dx}{\int_{\underline{v}-t}^{\overline{v}-t} f_\varepsilon(x)\,dx}\right)^2\right]$$

$$= \frac{\int_{\underline{v}-t}^{\overline{v}-t}\left(x - \frac{\int_{\underline{v}-t}^{\overline{v}-t} x f_\varepsilon(x)\,dx}{\int_{\underline{v}-t}^{\overline{v}-t} f_\varepsilon(x)\,dx}\right)^2 f_\varepsilon(x)\,dx}{\sigma_\varepsilon^2 \cdot \int_{\underline{v}-t}^{\overline{v}-t} f_\varepsilon(x)\,dx}.$$

Given $\underline{v} - t \le \frac{\int_{\underline{v}-t}^{\overline{v}-t} x f_\varepsilon(x) dx}{\int_{\underline{v}-t}^{\overline{v}-t} f_\varepsilon(x) dx} \le \overline{v} - t$, for any $x \in [\underline{v} - t, \overline{v} - t]$, we have

$$\left| x - \frac{\int_{\underline{v}-t}^{\overline{v}-t} x f_\varepsilon(x)\,dx}{\int_{\underline{v}-t}^{\overline{v}-t} f_\varepsilon(x)\,dx}\right| \le \overline{v} - \underline{v}\,.$$

Thus, $\boldsymbol{H}_{[\underline{v},\overline{v}]}(t) \le \frac{(\overline{v}-\underline{v})^2}{\sigma_\varepsilon^2}$.

A.8 The proof of Lemma 3

By the item (1) in appendix A.7, $\lim_{t\to+\infty} \boldsymbol{H}_{[\underline{v},\overline{v}]}(t) = 0$ and $\lim_{t\to-\infty} \boldsymbol{H}_{[\underline{v},\overline{v}]}(t) = 0$. So, there exists a positive constant $M$ such that $\boldsymbol{H}_{[\underline{v},\overline{v}]}(t) < 0.5$ holds for any t with $|t| > M$.

Since $\boldsymbol{H}_{[\underline{v},\overline{v}]}(t)$ is a continuous function of $t$, there exists a $t_0 \in [-M, M]$ such that $\boldsymbol{H}_{[\underline{v},\overline{v}]}(t)$ can reach its maximum in the interval $[-M, M]$. That is, for any $t \in [-M, M]$,

$$\boldsymbol{H}_{[\underline{v},\overline{v}]}(t) \le \boldsymbol{H}_{[\underline{v},\overline{v}]}(t_0).$$

By the item (3) in appendix A.7, $\boldsymbol{H}_{[\underline{v},\overline{v}]}(t_0) < 1$. In summary, for any $t \in R$, we derive

$$\boldsymbol{H}_{[\underline{v},\overline{v}]}(t) \le max\left\{\boldsymbol{H}_{[\underline{v},\overline{v}]}(t_0), 0.5\right\} < 1.$$

It follows that $\mathrm{Sup}_{t\in R}\, \boldsymbol{H}_{[\underline{v},\overline{v}]}(t) \le max\left\{\boldsymbol{H}_{[\underline{v},\overline{v}]}(t_0), 0.5\right\} < 1$.

By item (2) in appendix A.7, for any $t \in R$, $0 < \boldsymbol{H}_{[\underline{v},\overline{v}]}(t) < 1$, and then $\mathrm{Sup}_{t\in R}\, \boldsymbol{H}_{[\underline{v},\overline{v}]}(t) > 0$.

## Appendix